\documentclass[aps,apl,twocolumn,showpacs,groupedaddress]{revtex4}
\pdfoutput=1
\usepackage{dcolumn}   
\usepackage{bm}        
\usepackage{amssymb}   
\usepackage{verbatim}
\usepackage[pdftex]{graphicx}  

\usepackage{amsmath}
\usepackage{amsfonts}

\begin{document}

\title {Superconducting Low-Inductance Undulatory Galvanometer Microwave Amplifier}
\author{D. Hover, Y.-F. Chen, G. J. Ribeill, S. Zhu, S. Sendelbach, R. McDermott}
\email[Electronic address: ]{rfmcdermott@wisc.edu}

\affiliation{Department of Physics, University of
Wisconsin, Madison, Wisconsin 53706, USA}

\date{\today}

\begin{abstract}
We describe a microwave amplifier based on the Superconducting Low-inductance Undulatory Galvanometer  (SLUG). The SLUG is embedded in a microstrip resonator, and the signal current is injected directly into the device loop.  Measurements at 30 mK show gains of 25 dB at 3 GHz and 15 dB at 9 GHz. Amplifier performance is well described by a simple numerical model based on the Josephson junction phase dynamics. We expect optimized devices based on high critical current junctions to achieve gain greater than 15 dB, bandwidth of several hundred MHz, and added noise of order one quantum in the frequency range of 5-10 GHz.
\end{abstract}

\maketitle

\begin{figure}[t]
\begin{center}
\includegraphics[width=.45\textwidth]{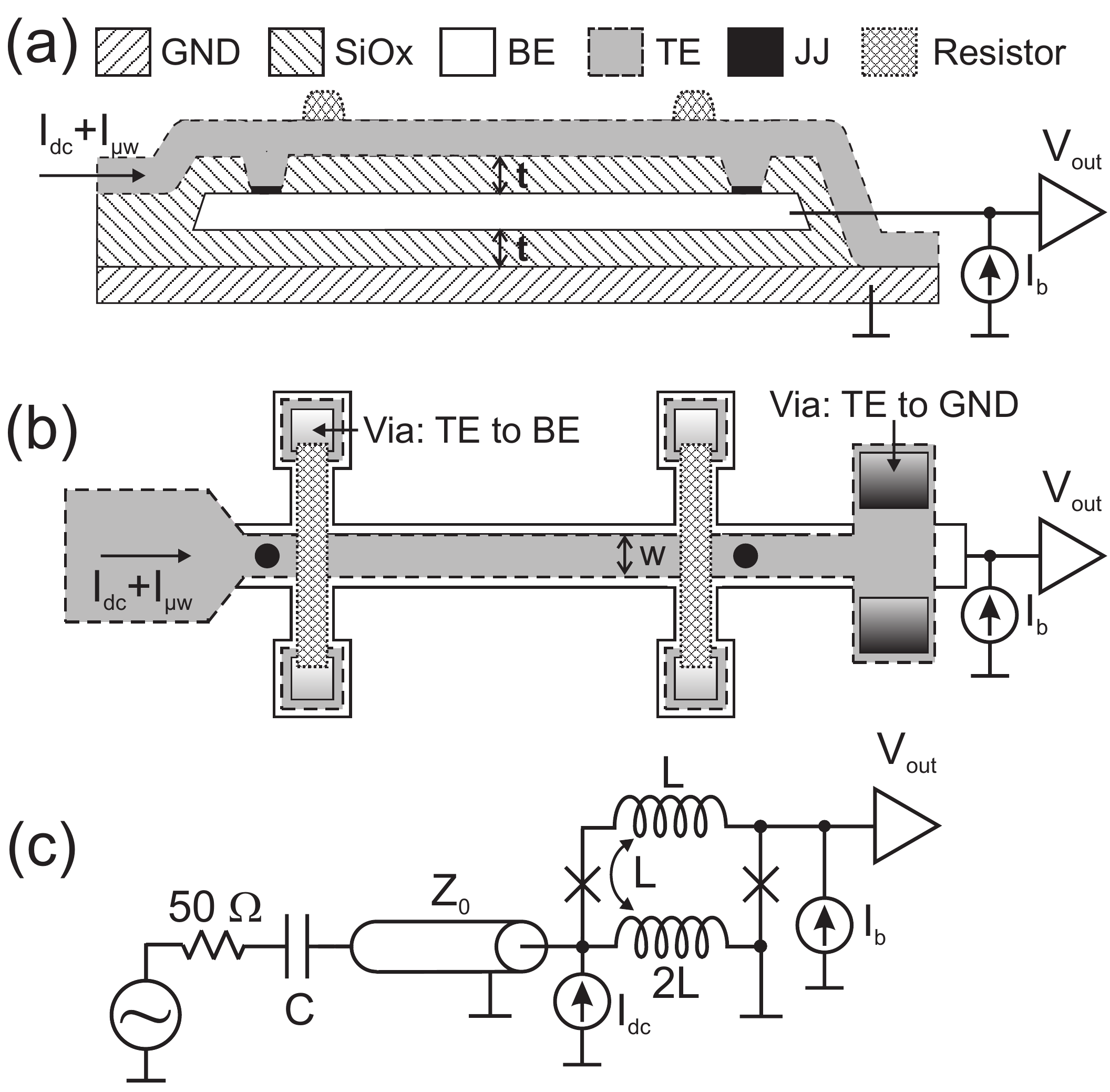}
\vspace*{-0.0in} 
\caption{SLUG microwave amplifier. (a) SLUG layer stackup. GND is the groundplane, JJ are the Josephson junctions, BE is the bottom electrode and TE is the top electrode. $I_b$ is the external bias current. $I_{dc}$ establishes the quasistatic flux bias point, while $I_{\mu w}$ is the microwave signal to be amplified. (b)  Layout of SLUG element, as seen from above (not to scale). (c) Schematic diagram of the SLUG microwave amplifier. A 50 $\Omega$ source is capacitively coupled to the SLUG via a microstrip transmission line with characteristic impedance $Z_0$ = 5.6 $\Omega$ and a bare quarter-wave resonance at 3.36 GHz. The transmission line is terminated in an inductive short to ground through the TE trace of the SLUG.}
\label{fig:figure1}
\end{center}
\end{figure}

Recent progress in the superconducting quantum circuit community has motivated a search for ultralow-noise microwave amplifiers for the readout of qubits and linear cavity resonators \cite{Schoelkopf04, Wallraff11}. It has long been recognized that the dc Superconducting QUantum Interference Device (SQUID) can achieve noise performance approaching the standard quantum limit of half a quantum \cite{Clarke81}. While in principle the SQUID should be able to amplify signals approaching the Josephson frequency (typically several tens of GHz), it remains challenging to integrate the SQUID into a 50 $\Omega$ environment and to provide for efficient coupling of the microwave signal to the device.  In one arrangement, a multiturn input coil has been configured as a microstrip resonator, with the SQUID washer acting as a groundplane \cite{Heiden98}. This so-called microstrip SQUID amplifier has achieved noise performance within a factor of two of the standard quantum limit at 600 MHz \cite{Clarke01, Clarke11}; however, performance degrades at higher frequencies due to the reduced coupling associated with the shorter microstrip input line \cite{Clarke03}. An alternative approach accesses the GHz regime by integrating a high-gain SQUID gradiometer into a coplanar transmission line resonator at a current antinode \cite{Aumentado08, Aumentado09}.  In other work, Jospehson circuits have been driven by an external microwave tone to enable ultralow-noise parametric amplification of microwave frequency signals \cite{Lehnert07, Tsai08, Siddiqi11}; however, Josephson parametric amplifiers provide limited bandwidth and dynamic range, and the external microwave bias circuitry introduces an additional layer of complexity. 

In this Letter we describe a new device configuration that provides efficient coupling of a GHz-frequency signal to a compact Superconducting Low-inductance Undulatory Galvanometer (SLUG). In contrast to the dc SQUID, which relies on a separate inductive element to transform the input signal into a magnetic flux, the SLUG samples the magnetic flux generated by a current that is directly injected into the device loop \cite{Clarke66}. The compact geometry of the SLUG makes the device straightforward to model at microwave frequencies and easy to integrate into a microwave transmission line.  Moreover, it is simple to decouple the SLUG modes from the input modes, allowing for separate optimization of the gain element and the matching network.

\begin{figure}[t]
\begin{center}
\includegraphics[width=.45\textwidth]{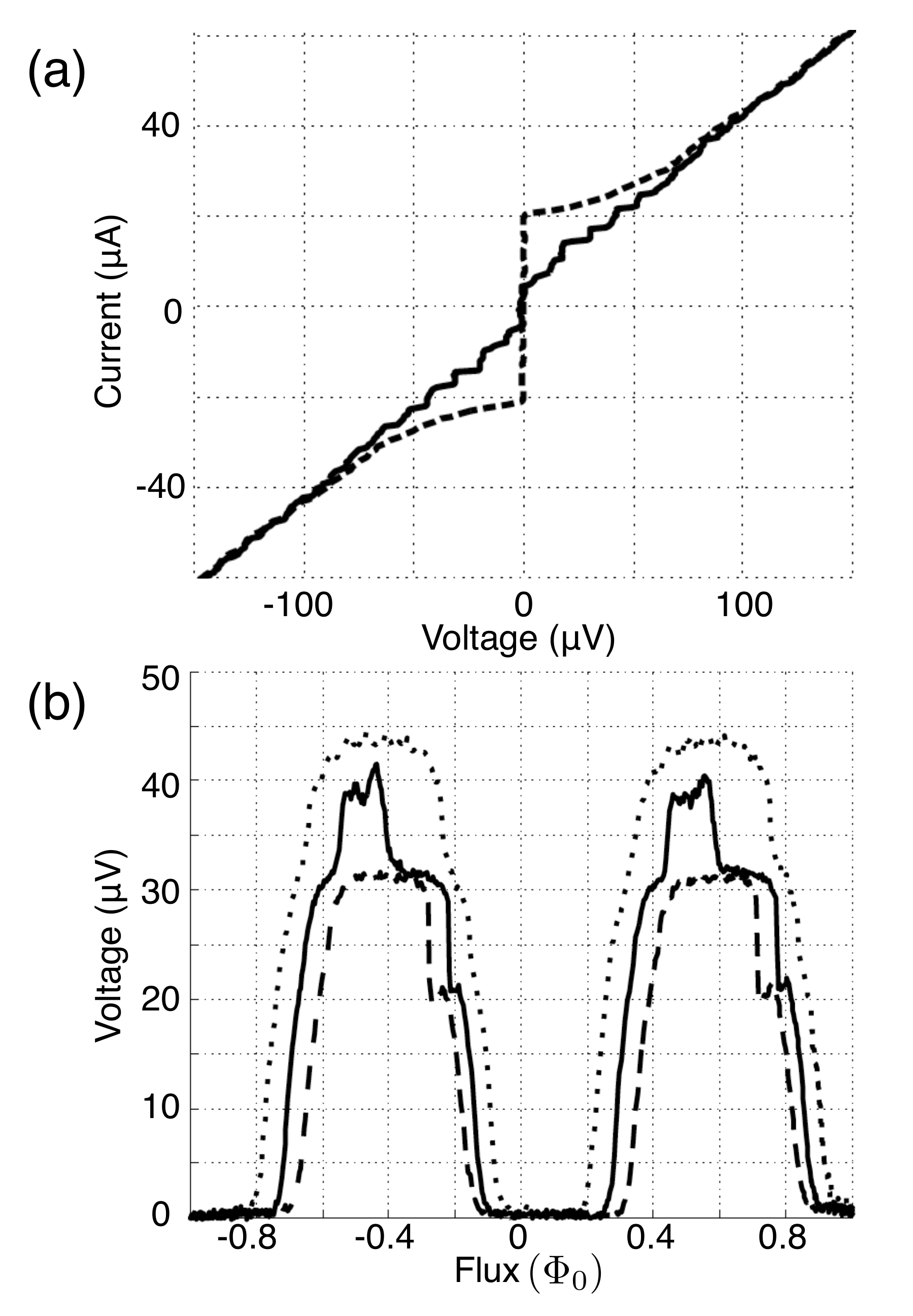}
\vspace*{-0.0in} \caption{DC characteristics of the SLUG measured at 30 mK.  (a) I-V curve of the device. The two traces correspond to a flux bias that maximizes (dash) and minimizes (solid) the critical current of the SLUG . (b) V-$\Phi$ curve of the device for three current biases: 16 $\mu$A (dash), 18 $\mu$A (solid) and 20 $\mu$A (dot).}
\label{fig:figure2}
\end{center}
\end{figure}

The layer stackup of the SLUG element is shown in Figure \ref{fig:figure1}(a). The amplifier is realized in three metallization steps corresponding to the circuit groundplane and the two arms of the SLUG loop; two dielectric layers separate the metal traces. The metal traces are formed from sputtered aluminum films defined by a wet etch, while the dielectric layers are formed from Plasma-Enhanced Chemical Vapor Deposition (PECVD) grown SiO$_2$ with vias defined by a CHF$_3$ dry etch. The AlO$_x$ barriers are formed by thermal oxidation at ambient temperature, and the nominal junction area is 2 $\mu$m$^2$. The normal metal shunt resistors are formed from evaporated palladium. A similar device geometry was studied in \cite{Clarke82}, although there was no additional groundplane layer and no attempt was made to integrate the SLUG element with a microwave transmission line. 

Figure \ref{fig:figure1}(b) shows the layout of the SLUG as seen from above. The SLUG loop inductance $L$ is determined from the self and mutual inductances of the base electrode (BE) and the top electrode (TE) traces: $L \approx L_{BE} +L_{TE} -2L_M$, where $L_{BE}$ ($L_{TE}$) is the self-inductance of the trace formed in the BE (TE) layer, and $L_M$ is the mutual inductance between the two traces. For a SLUG element with length $\ell$, trace width $w$, and with the BE (TE) trace separated from the groundplane by distance $t$ (2$t$), we find $L_{BE} \approx \mu_0 t \ell /w$, with $L_{TE} \approx 2 L_{BE}$ and $L_M \approx L_{BE}$. Therefore, we have $L \approx L_{BE}$. The current $I_b$ biases the device in the finite voltage state and the current $I_{dc}$ establishes a quasistatic flux bias point, while the current $I_{\mu w}$ is the microwave signal to be amplified. The currents $I_{dc}$ and $I_{\mu w}$ are combined by an on-chip bias network and injected into the TE trace of the SLUG, which is directly connected to the circuit groundplane. The mutual inductance $M$ of the injected current to the SLUG loop is approximately equal to the SLUG self-inductance $M \approx L$. 

A schematic diagram of the amplifier is shown in Figure \ref{fig:figure1}(c). The microwave signal from a 50 $\Omega$ source is coupled to the SLUG element via an on-chip capacitor $C$ = 150 fF and a  microstrip transmission line section formed in the TE and GND layers with characteristic impedance $Z_0$ = 5.6 $\Omega$ and a bare $\lambda/4$ resonance at 3.36~GHz. 

The I-V and V-$\Phi$ curves of the SLUG are shown in Figure \ref{fig:figure2}.  From the measured characteristics we infer an individual junction critical current $I_0$ = 10.8 $\mu$A, a junction shunt resistance $R$ = 5 $\Omega$, and a loop inductance $L \approx$ 13 pH. Therefore we find a dimensionless inductance $\beta_L \equiv 2I_0 L/\Phi_0$ = 0.14 and a damping parameter $\beta_C \equiv 2\pi I_0 C_j R^2/\Phi_0$ = 0.08, where $C_j$ = 100 fF is the junction self capacitance and $\Phi_0 = h/2e$ is the magnetic flux quantum. Sharp Shapiro step-like structure occurs at voltages corresponding to Josephson frequencies that are integer multiples of the half-wave resonance of the input circuit; this is a consequence of the strong coupling of the resonant input matching network to the SLUG element.

The microwave performance of the device was probed at 30~mK in a dilution refrigerator (DR). The dc bias currents $I_{dc}$ and $I_b$ were heavily filtered at 4.2 K and 30 mK, and the input microwave signal was coupled to the amplifier with 40 dB of cold attenuation.  The output signal was coupled through a commercial bias-T to a low-noise HEMT amplifier at 4.2 K.  Low-loss cryogenic coax relay switches were mounted on the mixing chamber plate of the DR to perform an $\textit{in situ}$ calibration of amplifier gain.

\begin{figure}[t]
\begin{center}
\includegraphics[width=.45\textwidth]{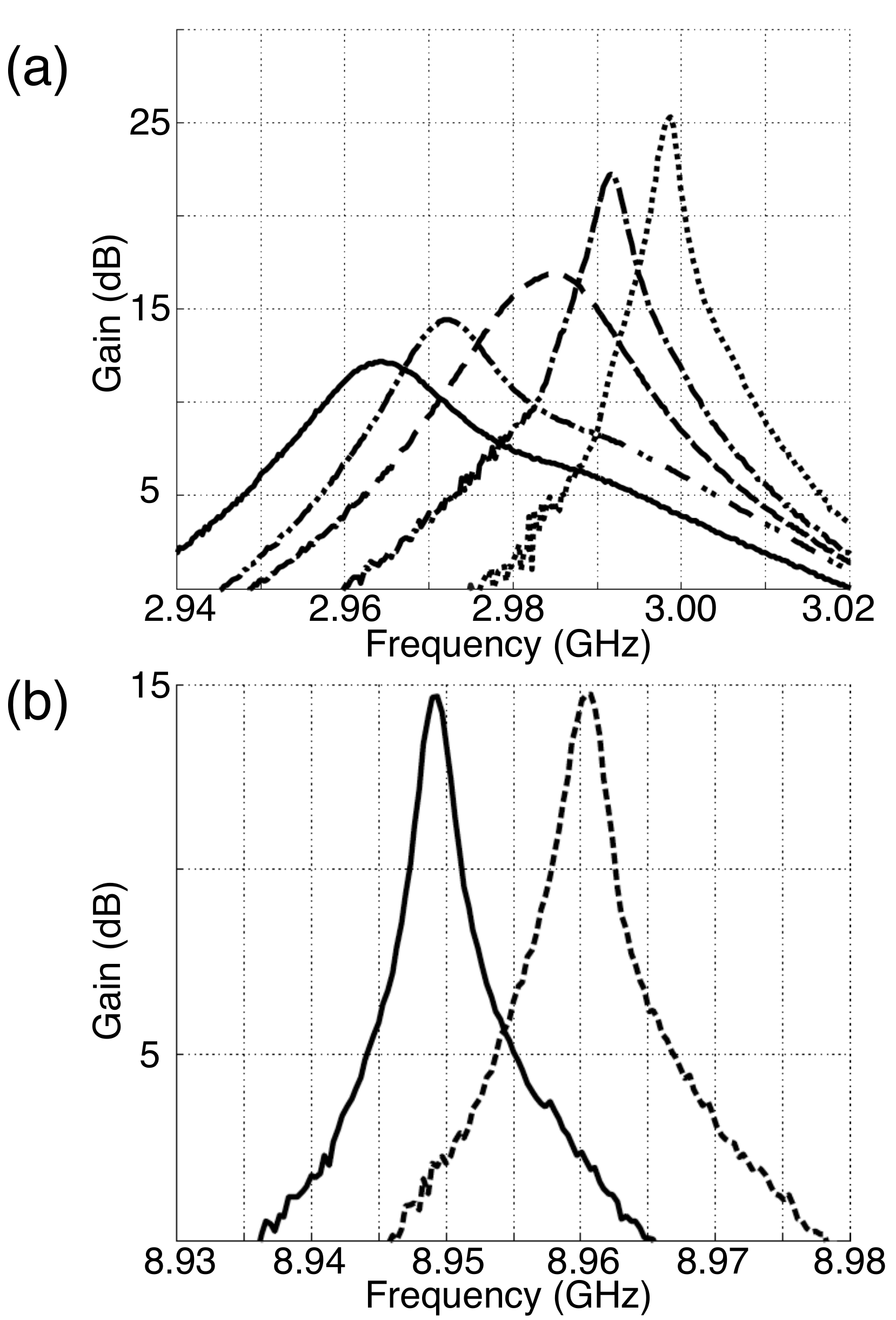}
\vspace*{-0.0in} \caption{Forward gain of the SLUG amplifier. (a) Gain at the first resonant mode for $I_b$ = 20 $\mu$A and various flux bias points. (b) Amplifier gain at the second resonant mode for two different bias points.}
\label{fig:figure3}
\end{center}
\end{figure}

Figure \ref{fig:figure3}(a) shows gain curves for different flux biases at a SLUG current bias $I_b$ = 20 $\mu$A. A  maximum gain of 25 dB is achieved at an operating frequency close to 3 GHz with a bandwidth of several MHz. The measured amplifier performance agrees with a theoretical treatment of the device using the method outlined in \cite{McDermott11}.  The compact SLUG loop, with a sensing area on the order of 30 $\mu$m$^2$, makes the device insensitive to environmental magnetic fluctuations: the amplifier could be left biased at a high gain point for several hours without any noticeable degradation in the frequency dependent gain of the device.  

It is possible to achieve substantial gain at significantly higher operating frequencies by driving the amplifier at a higher-order harmonic of the input circuit. In Figure \ref{fig:figure3}(b) we show the frequency dependent gain of the SLUG amplifier at an operating frequency near 9 GHz, corresponding to the $3\lambda/4$ resonant mode of the input matching network. 

The gain-bandwidth tradeoff apparent from Figure \ref{fig:figure3}(a) can be understood from a simple model. The maximum power gain $G_m$ available from the SLUG element can be expressed as follows: 

\begin{align}
G_m = \frac{M^2 V_\Phi^2}{4R_i R_o} =  \frac{1}{4\rho_i \rho_o}\left(\frac{V_\Phi}{\omega}\right)^2
\label{eq:powergain}
\end{align}

\noindent Where $V_\Phi=\frac{\partial V}{\partial \Phi}$ is the flux-to-voltage transfer function, $\omega$ is the angular frequency of the amplified signal, and $\rho_{i,o}$ are bias-dependent (and frequency-independent) dimensionless impedance parameters related to the input and output impedances $R_{i,o}$ as follows: $R_i = \rho_i \frac{\omega^2 M^2}{R}$ and $R_o = \rho_o R$.  The analysis of \cite{McDermott11} shows that for $I_b<2I_0$ and for a narrow range of flux biases, $\rho_i$ becomes vanishingly small and power gain becomes large. On the other hand, amplifier bandwidth scales as $R_i^{1/2}$.  It follows that even an unoptimized device will have impressive gain at a few bias points where $\rho_i$ approaches zero; however, amplifier bandwidth will be quite modest for these bias points. Likewise, for flux bias points that access higher $\rho_i$, amplifier gain decreases while bandwidth increases. The bandwidth for the device described above is also limited by weak coupling to the source through the small coupling capacitance $C$. 

It is important to note that the amplifier described in this Letter is far from optimized. In particular, a larger critical current density will enhance $V_\Phi$ significantly and make it possible to achieve reasonable gain while maintaining a large bandwidth in excess of 100~MHz.  For example, a high critical current Nb-AlO$_x$-Nb junction process could provide a factor of ten increase in $I_0$, resulting in an improvement in gain approaching 20~dB. A rigorous numerical analysis confirms this simple picture, where simulations of a SLUG with $L$ = 10 pH, $\beta_L = 1$, and $\beta_C = 0.8$ yield gain of approximately 15~dB at 5 GHz, bandwidth of order several hundred MHz, and less than one quantum of added noise. Additionally, the Shapiro step-like structure can be reduced by decoupling the SLUG modes from the input modes via a filter inductance; this will improve the dynamic range of the amplifier and simplify integration of the SLUG element into a more complicated microwave environment \cite{McDermott11}. 

To conclude, we have realized a SLUG-based microwave amplifier with gains of 25 dB at 3 GHz and 15 dB at 9 GHz and with bandwidth of several MHz. The measured microwave performance agrees well with the numerical model of \cite{McDermott11}. We expect optimized SLUG amplifiers to achieve gain in excess of 15 dB, bandwidth of several hundred MHz and added noise of the order one quantum. Low-noise, broadband SLUG microwave amplifiers could play an enabling role in single-shot dispersive qubit readout \cite{Schoelkopf04}, dark-matter axion detection \cite{Clarke10} or fundamental studies of microwave photon statistics \cite{Wallraff11} and microwave emission from Josephson junctions and other mesoscopic samples \cite{Esteve11}.

\begin{acknowledgments}
This work was supported by the DARPA/MTO QuEST program through a grant from AFOSR. Additional support came from IARPA under contracts W911NF-10-1-0324 and W911NF-10-1-0334. All statements of fact, opinion or conclusions contained herein are those of the authors and should not be construed as representing the official views or policies of the U.S. Government.
\end{acknowledgments}

\end{document}